\documentclass[11pt]{article}
\RequirePackage[OT1]{fontenc}
\RequirePackage{amsthm,amsmath}
\RequirePackage[numbers]{natbib}

\numberwithin{equation}{section}
\theoremstyle{plain}
\newtheorem{thm}{Theorem}[section]

\usepackage{graphicx} %
\usepackage{times}                  
\usepackage{tabularx}

\newtheorem{lem}[thm]{Lemma}
\newtheorem{rem}[thm]{Remark}
\newtheorem{defn}[thm]{Definition}
\newcommand{\ad}{&\!\!\!\disp}
\newcommand{\aad}{&\disp}
\newcommand{\barray}{\begin{array}{ll}}   
\newcommand{\earray}{\end{array}}

\usepackage{makeidx}  
\usepackage{amsfonts}
\makeindex            
\usepackage{times}  
\usepackage{graphicx} 

\newcommand{\rr}{{\Bbb R}}

\newcommand{\one}{{\hbox{1{\kern -0.35em}1}}}

\newcommand{\M}{{\cal M}}

\newcommand{\beq}[1]{\begin{eqnarray} \label{#1}}
\newcommand{\eeq}{\end{eqnarray}}
\newcommand{\bed}{\begin{displaymath}}
\newcommand{\eed}{\end{displaymath}}
\newcommand{\bea}{\bed\begin{array}{rl}}
\newcommand{\eea}{\end{array}\eed}
\newcommand{\disp}{\displaystyle}
\newcommand{\al}{\alpha}

\newcommand{\thmref}[1]{Theorem~{\rm \ref{#1}}}
\newcommand{\lemref}[1]{Lemma~{\rm \ref{#1}}}

\def\openbox{$\sqcup\llap{$\sqcap$}$}
\def\endproof{\unskip \enskip \null \nobreak \hfill \openbox \par}
\begin{document}

\title{Optimal Resource Extraction in Regime Switching L\'{e}vy Markets}
\author{Moustapha Pemy\thanks{Department of Mathematics, Towson
University, Towson, MD 21252-0001, mpemy@towson.edu
 } 
 }
\date{}
\maketitle


\begin{abstract} 
This paper studies the problem of optimally extracting nonrenewable natural resources. Taking into account the fact that the market values of the main natural resources such as oil, natural gas, copper,...,etc, fluctuate randomly following global and seasonal macroeconomic parameters, these values are modeled using Markov switching L\'evy processes. We formulate this problem as a finite time horizon combined optimal stopping and optimal control problem. We characterize the value function of this problem as the unique viscosity solution of the corresponding Hamilton-Jacobi-Bellman equations. Moreover, we propose a finite difference approximation of the value function and prove its convergence. This enables us to derive optimal extraction and stopping policies. A numerical example is presented to illustrate these results.
\end{abstract}




\vspace*{0.2in}

\section{Introduction}

The optimal extraction of nonrenewable natural resources has received a great deal of interest in the literature 
since the early thirties. The first major contribution to this problem was made by  Hotelling (1931), he proposed an extraction model  in which the commodity price is deterministic and was able to derive an optimal extraction policy. Many economists have 
extended the Hotelling model by taking into account
the uncertainty in the supply and the demand of strategic commodities. Among many others, one can cite the work of  Hanson (1980), Solow and Wan (1976), Pindyck (1978) and (1980), Sweeney (1977),  Lin and  Wagner (2007) for various extensions of the basic Hotelling model.\\
It is self-evident that prices of commodities such as oil, natural gas, copper, and gold are greatly uncertain and fluctuate following divers macroeconomic and global  geopolitical forces.
It is, therefore, crucial to take into account the random dynamic of the commodity value when solving the optimal extraction problem in order to maintain  the validity of the result obtained in the procedure. In this paper, we use  regime switching L\'{e}vy processes to model the prices of natural resources. These processes will help us capture both the seasonality and spikes frequently observed in the market prices of natural resources such as oil and natural gas. Moreover, given that the vast majority of mining leases have finite time horizons, and that there is always the possibility that before the end of a given lease the extraction cost and the change in the commodity value may force the mining company to stop extracting. We, therefore, treat this optimal extraction problem as a combined optimal control and optimal stopping problem.    
Optimal stopping and control problems over finite time and infinite time horizons
 have generated a good deal of interest in the literature, various 
applications have been developed in many areas of science, engineering, and finance. A wide range of techniques have been used to tackle these
 problems. Early treatment of optimal stopping of Markov processes can be 
traced back to Dynkin (1963), Snell (1952) and McKean (1960). 
Since the early eighties, the method of viscosity solutions  introduced by 
Crandall and Lions (1992)  has been widely used  to solve optimal 
control and optimal stopping problems. Many authors have used the viscosity
 solutions machinery to solve optimal stopping and/or control  of It\^{o} diffusions, 
one can refer to  Soner (1986),  $\O$ksendal and Reikvam (2004), Pemy (2005), (2011) and (2014)
 among others. For more on the theory and application of viscosity solutions 
on can refer to Crandall, Ishii and Lions (1992), Fleming and Soner 
(2006), Yong and Zhou (1999).\\
As we all know,  the prices of natural resources such as energy commodities usually feature various spikes and shocks, due to political instabilities in producing countries and the growing global demand for energy. We use L\'{e}vy processes coupled with a hidden Markov chain to capture both  jumps and seasonality in the commodity prices. L\'{e}vy processes and jump diffusions have also been widely studied in the
 literature. The optimal control and optimal stopping of these processes have
 been investigated by many authors, one can refer to $\O$ksendal and 
Sulem (2004), Hanson (1980),  Pham (1998) and Pemy (2014). Roughly speaking, regime switching L\'{e}vy processes consist 
of L\'{e}vy processes with an additional source of randomness, namely,
 a hidden Markov process $(\al(t))_t$ in continuous time or $(\al_n)_n$
 in discrete time. The process $(\al(t))_t$ is the finite state Markov 
chain, it captures the different changes in regime of the L\'{e}vy process.
 Regime switching modeling has been widely used in the many fields since
 its introduction by Hamilton (1989) in time series analysis.  Many 
authors have studied the control of systems that involve regime switching
 using a hidden Markov chain, one can cite Zhang and Yin (1998), (2005), Pemy and Zhang (2006)
 among others. \\ 
In this paper, we treat the problem of finding optimal strategies for extracting a natural resource as a combined optimal control and stopping  of Markov switching L\'{e}vy processes in a finite time horizon. The main contribution of this paper is two-fold, first, we prove that the value function is the  unique viscosity solution of the associated 
Hamilton-Jacobi-Bellman equations. Then, we build a finite difference approximation scheme  and prove its convergence to the unique viscosity solution of HJB equations. This enables us to derive both the optimal extraction policy and the stopping policy.\\ 
The paper is organized as follows. In the next section, we
formulate the problem under consideration.
In Section 3, we obtain the continuity property of the value function
and show that it is the unique  viscosity solution of the HJB equations, moreover, we derive both the optimal extraction  and stopping policies. And in section 4, we construct  a finite difference approximation scheme and prove its convergence to the value function. Finally, in section 5, we give a numerical example.

\section{Problem formulation}
Consider a company that has a mining lease with expiration $0<T<\infty$.  Let $m$ be an integer $m\geq2$, and  $\al(t)\in {\cal M} = \{1,2,...,m\}$ be a Markov chain
with generator $Q=(q_{ij})_{m,m}$, i.e., $q_{ij}\geq 0$ for $i\neq j$,  and $\Sigma_{j=1}^m q_{ij}=0$ for $i \in {\cal M}$. In fact, the Markov chain $(\al(t))_t$ will capture various states of the commodity market.  Let $(\eta_t)_t$ be a L\'{e}vy process, and let $N$ be the Poisson random measure of $(\eta_t)_t$,  for any Borel set $B\subset \rr$, 
$
\disp N(t,B)=\sum_{0<s\leq t}{\bf 1}_B(\eta_s-\eta_{s^-}). 
$
The differential form of $N$ is denoted by $N(dt,dz)$. Let $\nu$ be the L\'{e}vy measure of $(\eta_t)_t$, we have $\nu(B)=E[N(1,B)]$ for any Borel set $B\subset \rr$. 
We define the differential form $\bar{N}(dt,dz)$ as follows,
\bea 
\disp
\bar{N}(dt,dz)=\left\{\begin{array}{ll} N(dt,dz)-\nu(dz)dt \qquad &\hbox{if  } |z|<1\\
   N(dt,dz)&\hbox{if  } |z|\geq 1.   \end{array} \right.
   \eea
From L\'evy-Khintchine formula we have,
\beq{opep}
\disp \int_{\rr}\min(|z|^2,1)\nu(dz)<\infty.
\eeq
We assume that  the L\'{e}vy measure $\nu$ has finite intensity,
\beq{finiteIntensity} 
\disp \Gamma = \int_{\rr} \nu(dz)<\infty.
\eeq
 In other words, the total sum of jumps and spikes of the commodity price during the lifetime of the contract is finite.
Let $X(t)$ denote the price of one unit of a  natural resource  at time $t$. And let $Y(t)$ represent the size of remaining resources at time $t$. 
We assume that our extraction activity can be modeled by the process $u(t)$ taking values in the interval $U=[0,K]$, $u$ is in fact the extraction rate of the resource in question.  The processes $X(t)$ and $Y(t)$ satisfy the following stochastic differential equations. 
\beq{defi3}
\left \{ \begin{array}{ll} \disp\mathrm{d}X(t)= \bigg(\mu(t, X(t),u(t),\alpha(t))\mathrm{d}t+\sigma(t,X(t),u(t),\alpha(t))\mathrm{d}W(t)  \\
 \disp \hspace{1 in} + \int_{\rr}\gamma(t,X(t),u(t),\alpha(t),z)\bar{N}(dt,dz)\bigg),\\
 dY(t)= -u(t) dt, \\
\disp X(s)=x, \,\,\,\, Y(s) = y \geq 0, \qquad s\leq t\leq T. \end{array} \right.
\eeq   
where $x$ and $y$ are the initial values, $T$ is a finite time. $W(t)$ is the standard Wiener process on $\rr$,
we assume that $(W(t))_t$, $(\eta_t)_t$ and $(\alpha(t))_t$ are defined on a probability space 
$(\Omega,{\cal F },P)$, and are independent.  The process $u(t)$ is referred as the control process in this model. 

\begin{rem}
Our commodity pricing model (\ref{defi3}) encompasses a wide range of possibilities. Below are some of the particular cases of our general model.
\begin{enumerate}
 \item If the size of the mine is not large enough to influence the price of the commodity then the parameters $\mu$, $\sigma$ and $\gamma$ will not depend on the extraction activities of the mining company. A typical example, in this case, is the exponential L\'{e}vy model for commodity prices 
\beq{ExpoLevy}
\disp\mathrm{d}X(t)= X(t)\bigg(\mu(\alpha(t))\mathrm{d}t+\sigma(\alpha(t))\mathrm{d}W(t)  + \gamma(\alpha(t))\int_{\rr}z\bar{N}(dt,dz)\bigg).
\eeq
This model is appropriate for most mining problems as well as derivative pricing problems.
\item If the size of the mine is large enough or the country where the mine is located is one of the major producers of the commodity in question such as the Saudi Arabia is for oil, then the extraction policies of such a country will definitely affect the world price of the commodity. In this case, we can assume that the drift of the price process will depend on the extraction rate. However, one can foresee  a case where even the diffusion and the jump coefficients are also influenced by the extraction rate. The typical pricing model, in this case, has the form
\beq{ExpoMain}
\disp\mathrm{d}X(t)= X(t)\bigg((\mu(\alpha(t))-\lambda u(t))\mathrm{d}t+\sigma(\alpha(t))\mathrm{d}W(t)  + \gamma(\alpha(t))\int_{\rr}z\bar{N}(dt,dz)\bigg),
\eeq    
where $\lambda\in(0,1)$ captures the relative impact of the extracting activities.
\end{enumerate}
In sum, we will study this interesting problem  in its more generalized form as stated in (\ref{defi3}).
\end{rem}
The functions $\mu: [0,T]\times\rr\times[0,K]\times{\cal M}\rightarrow \rr$,  $\sigma: [0,T]\times\rr\times[0,K]\times{\cal M}\rightarrow \rr$ and $\gamma:[0,T]\times\rr\times[0,K]\times{\cal M}\times\rr\rightarrow\rr$  satisfy the following properties:
\begin{itemize}
\item {\bf Lipschitz continuity: }  There exists a constant $C>0$ such that  
\beq{condition1}
|\mu(t,x,v,i)-\mu(t,y,v,i)|^2+|\sigma(t,x,v,i)-\sigma(t,y,v,i)|^2+\int_{|z|<1}|\gamma(t,x,v,i,z)\nonumber \\
-\gamma(t,y,v,i,z)|^2\nu(dz)<C|x-y|^2, \quad \mbox{for all } t, x, y, v.
\eeq
\item {\bf Growth condition:} There exists a constant $C>0$ such that
\beq{condition2}
 |\mu(t,x,v,i)|^2+|\sigma(t,x,v,i)|^2+\int_{|z|<1}|\gamma(t,x,v,i,z)|^2\nu(dz)<C(1+|x|^2), \\
\quad \mbox{for all } t, x, y, v.  \nonumber
\eeq
\end{itemize} 
The assumptions (\ref{condition1}) and (\ref{condition2}) guarantee that for any Lebesgue measurable  control $u(\cdot)$ on $[s,T]$, the equation (\ref{defi3}) has a unique solution. For more one can refer to  $\O$ksendal and Sulem (2004). For each initial data $(s,x,y,i)$ we denote by ${\cal U}(s,x,y,i)$ the set of admissible controls which is just the set of all controls $u(\cdot)$ that are $\{{\cal F}_t\}_{t\geq 0}$-adapted where $ {\cal F}_t = \sigma\{\alpha(\xi),W(\xi); \xi\leq t\}$ and such that the equation (\ref{defi3}) has a solution with initial data  $X(s)=x$, $Y(s) =y$, $\al(s) = i$.\\ 
Let 
$\Lambda_{s,T}$ denote the set of ${\cal F}_t$-stopping times $\tau$ such that  $s\leq\tau\leq T$ almost surely.  Given a discounting factor $r> 0$, we define the payoff functional as follows
\beq{payoff}
&&J(s,x,y,i;u,\tau)\nonumber \\
&=&E\bigg[\int_s^\tau e^{-r(t-s)} L(t,X(t),Y(t),u(t),\al(t))dt \nonumber \\ 
.  && + e^{-r(\tau-s)}\Phi(\tau,X(\tau),Y(\tau), \al(\tau))\bigg{| } X(s)=x,Y(s) = y, \al(s)=i\bigg].
\eeq
For each $i \in {\cal M}$, the functions $L(t,x,y,u,i)$ and $\Phi(t,x,y,i)$ are continuous with respect to their arguments $t$ and $u$ and are Lipschitz continuous with respect to the arguments $x$ and $y$. For simplicity, we will use ${\cal U} $ to denote  $ {\cal U}(s,x,y,i)$ and similarly we will use $\Lambda $ to denote $\Lambda_{s,T}$. Our goal is to find the control $u^*\in {\cal U}(s,x,y,i)$ and a stopping time $\tau^* \in \Lambda_{s,T}$ such  that 
\beq{val2}
V(s,x,y,i)=\sup_{u \in {\cal U}, \tau \in \Lambda }J(s,x,i;u, \tau)=J(s,x,y,i; u^*,\tau^*).
\eeq
 The function $V(s,x,y,i)$ is called the value function of the combined optimal stopping and control problem.
The process $(X(t),Y(t),\al(t))$ is a Markov process with generator ${\cal L}^w$,   defined as follows
\beq{genra}
\disp
({\cal L}^w f)(s,x,y,k)&=&\frac{\partial f(s,x,y,k)}{\partial s}+\frac{1}{2}\sigma^2(s,x,w,k)\frac{\partial^2f(s,x,y,k)}{\partial x^2}  \nonumber\\
&&+  \mu(s,x,w,k)\frac{\partial f(s,x,y,k)}{\partial x}
 +\int_{\rr}\bigg(f(s,x+\gamma(s,x,w,k,z),y,k)\nonumber\\
&&-f(s,x,y,k) -{\bf 1}_{\{|z|<1\}}(z)\frac{\partial f(s,x,y,k)}{\partial x}\cdot \gamma(s,x,w,k,z)\bigg)\nu(dz) 
\disp  \nonumber\\
&&-  w\frac{\partial f(s,x,y,k)}{\partial y} +  Qf(s,x,y,\cdot)(k), 
\eeq
for all  $s\in [0,T], x\in \rr, y\in \rr^+, w\in U, k\in{\cal M}, f(\cdot,\cdot,k)\in C^{1,2,1}_0([0,T]\times\rr\times\rr^+)$
with
\beq{Qgen}
\disp
 Qf(s,x,y,\cdot)(i)=\sum_{j\not =i}q_{ij}(f(s,x,y,j)-f(s,x,y,i)), 
\eeq
the generator of the Markov chain  $\al_t$.
In order to simplify the notation, we define the operator ${\cal G}$ as follows
\beq{GHamiltonian}
&&{\cal  G} (s,x,y,i,V(\cdot),V_s(\cdot), V_x(\cdot),V_y(\cdot),V_{xx}(\cdot)) \\
&=& rV(s,x,y,i)-\sup_{u\in U} \Bigg{(}\frac{\partial V(s,x,y,i)}{\partial s}+\frac{1}{2}\sigma^2(s,x,u,i)\frac{\partial^2V(s,x,y,i)}{\partial x^2}  \nonumber\\
&&+  \mu(s,x,u,i)\frac{\partial V(s,x,y,i)}{\partial x}
 +\int_{\rr}\bigg(V(s,x+\gamma(s,x,u,i,z),y,i)\nonumber\\
&&-V(s,x,y,i) -{\bf 1}_{\{|z|<1\}}(z)\frac{\partial V(s,x,y,i)}{\partial x}\cdot \gamma(s,x,u,i,z)\bigg)\nu(dz) 
\disp  \nonumber\\
&&-  u\frac{\partial V(s,x,y,i)}{\partial y} +  QV(s,x,y,\cdot)(i) + L(s,x,y,u,i)\Bigg{)}, \nonumber
\eeq 
the Hamiltonian of the system is given by 
\beq{hamiltonian}
& &{\cal  H} (s,x,y,i,V(\cdot),V_s(\cdot), V_x(\cdot),V_y(\cdot),V_{xx}(\cdot))  \\
&=&  \min \Bigg{[}{\cal  G} (s,x,y,i,V(\cdot),V_s(\cdot), V_x(\cdot),V_y(\cdot),V_{xx}(\cdot)) ,V(s,x,y,i)-\Phi(s,x,y,i) \Bigg{]}.\nonumber
\eeq
It is well known that the value function  $V(s,x,y,i) $  must formally satisfy the following HJB equation
\beq{sys}
\quad \left \{\begin{array}{ll}
\disp{\cal  H} (s,x,y,i,V(s,x,y,i),V_s(s,x,y,i), V_x(s,x,y,i),V_y(s,x,y,i),V_{xx}(s,x,y,i))=0, \\
 \hspace{1 in}  \hbox{for}\,\,(s,x,y,i)\in [0,T)\times \rr\times \rr^+\times {\cal M}, \label{Hamilton}\\
V(T,x,y,\alpha(T))=\Phi(T,x,y,\al(T)).  &\hspace{0.1 in} \end{array}\right. 
\eeq
Equation (\ref{sys}) is a fully nonlinear system of integro-differential equations, we may not have smooth solutions. So we will look for a weaker form of solution for this system, namely, viscosity solutions introduced by  introduced by Crandall and Lions (1983). Let us first recall the definition of viscosity solutions.  
\begin{defn}
Let $f:[0,T]\times\rr \times \rr^+ \times {\cal M} \rightarrow \rr$ such that $f(T,x,y,\alpha(T))=\Phi(T,x,y,\al(T))$ and for each $\iota \in {\cal M}$, $f(\cdot,\cdot,\cdot,\iota) \in \mathnormal{C}([0,T]\times \rr\times\rr^+)$. 
\begin{enumerate}
\item $f$ is a viscosity subsolution of the system (\ref{sys}) if for each $\iota \in {\cal M}$,
\beq{max}
{\cal H}\bigg(s_0,x_0,y_0,\iota, f(s_0,x_0,y_0, \iota), \frac{\partial \phi(s_0,x_0,y_0)}{\partial s  },   \frac{\partial \phi(s_0,x_0,y_0)}{\partial x  }, \nonumber\\
 \frac{\partial \phi(s_0,x_0,y_0)}{\partial y  },  \frac{\partial^2 \phi(s_0,x_0,y_0)}{\partial x^2  }\bigg) 
    \leq 0 \nonumber \\ 
\eeq
whenever $ \phi(s,x,y) \in \mathnormal{C}^{1,2,1}([0,T]\times\rr\times \rr^+)$ such that $f(s,x,y,\iota)-\phi(s,x,y)$ has local maximum at $(s,x,y)=(s_0,x_0,y_0)$.
\item $f$ is a viscosity supersolution of the system (\ref{sys}) if for each $i \in {\cal M}$,
\beq{min}
{\cal H}\bigg(s_0,x_0,y_0,\iota, f(s_0,x_0,y_0, \iota), \frac{\partial \psi(s_0,x_0,y_0)}{\partial s  },   \frac{\partial \psi(s_0,x_0,y_0)}{\partial x  }, \nonumber \\
 \frac{\partial \psi(s_0,x_0,y_0)}{\partial y  },  \frac{\partial^2 \psi(s_0,x_0,y_0)}{\partial x^2  }\bigg) \nonumber     \geq 0 \nonumber \\ 
\eeq
whenever $ \psi(s,x,y) \in \mathnormal{C}^{1,2,1}([0,T]\times\rr\times\rr^+)$ such that $f(s,x,y,\iota)-\psi(s,x,y)$ has local minimum at $(s,x,y)=(s_0,x_0,y_0)$.
\item  $f$ is a viscosity solution of (\ref{sys}) if it is both a viscosity subsolution and supersolution of (\ref{sys}).
\end{enumerate} 
\end{defn}
\section{Characterization of the value function}
In this section, we study the continuity of the value function and 
show  it is the unique viscosity solution of the HJB equation (\ref{sys}).
We first show the continuity property.
\begin{lem}\label{conti}
For each $i\in{\cal M}$,
the value function $V(s,x,y,i)$ is continuous in $(s,x,y)$.
Moreover, it has at most linear growth rate, i.e.,
there exists a constant $C$ such that
$\mid V(s,x,y,i) \mid \leq C(1+|x|+|y|)$.
\end{lem} 
The continuity of the value function with respect to the arguments $s, x$ and $y$ comes naturally from the application of the It\^{o}-L\'{e}vy isometry, the Lipschitz continuity assumption (\ref{condition1}) and the Gronwall's inequality.
\paragraph{Proof.}
Let $i\in{\cal M}$, we will first prove the  continuity in $x$. Let $(s,x_1,y,i)$ and $(s,x_2,y, i)$ be two initial data such  that we have two solutions $(X_1,Y)$ and $(X_2,Y)$ with $X_1(s)=x_1$, $X_2(s)=x_2$ and $Y=y$. Without loss of generality, we set ${\cal U }\equiv {\cal U}(s,x_1,y,i)\cap {\cal U}(s,x_2,y,i)$. We will use the notation $E^{s,x_1,x_2,y,i}[\,\, \cdot\,\,]$ to represent $E[\,\, \cdot\,\,|X_1(s)=x_1,X_2(s)=x_2,Y(s) = y, \al(s)=i]$. \\ It is clear that  ${\cal U}$  is non empty, thus for any control $u \in  {\cal U}$ and $t\in [s,T]$,  we have
\beq{ruur} 
\disp
&&X_1(t)-X_2(t) \nonumber \\
&=&x_1-x_2+\int_s^t[\mu(t,X_1(\xi),u(\xi),\al(\xi))-\mu(t,X_2(\xi),u(\xi),\al(\xi))]d\xi\nonumber \\
&& +\int_s^t[\sigma(t,X_1(\xi),u(\xi),\al(\xi))-\sigma(t,X_2(\xi),u(\xi),\al(\xi))]dW(\xi) \nonumber \\
 &&+\int_s^t\int_{\rr}[\gamma(t,X_1(\xi),u(\xi),\al(\xi),z)-\gamma(t,X_1(\xi),u(\xi),\al(\xi),z)]\bar{N}(d\xi,dz).
\eeq
Using the It\^{o}-L\'{e}vy isometry and the Lipschitz continuity assumption (\ref{condition1}),  we have
\beq{ruur} 
\disp
E^{s,x_1,x_2,y,i}(X_1(t)-X_2(t))^2&\leq &C_0|x_1-x_2|^2+C_1\int_s^tE^{s,x_1,x_2,y,i}(X_1(\xi)-X_2(\xi))^2d\xi\nonumber \\
 & &+C_2\int_s^tE^{s,x_1,x_2,y,i}(X_1(\xi)-X_2(\xi))^2d\xi \nonumber \\
 & &+C_3\int_s^t\int_{|z|<1}E^{s,x_1,x_2,y,i}|\gamma(t,X_1(\xi),u(\xi),\al(\xi),z)\nonumber \\
 & &\hspace{1.2 in}-\gamma(t,X_1(\xi),u(\xi),\al(\xi),z)|^2\nu(dz)d\xi \nonumber \\
 &\leq&C_0|x_1-x_2|^2+\max(C_1,C_2)\int_s^tE^{s,x_1,x_2,y,i}(X_1(\xi)-X_2(\xi))^2d\xi\nonumber \\
 & &+C_3\int_s^tE^{s,x_1,x_2,y,i}(X_1(\xi)-X_2(\xi))^2d\xi.
\eeq
Let $C=\max(C_0,C_1,C_2,C_3)$, (\ref{ruur}) becomes
\beq{ruur1} 
\disp
E^{s,x_1,x_2,y,i}(X_1(t)-X_2(t))^2
  &\leq&C|x_1-x_2|^2+C\int_s^tE^{s,x_1,x_2,y,i}(X_1(\xi)-X_2(\xi))^2d\xi.
\eeq
Applying the Gronwall's inequality, we have
\[ 
E^{s,x_1,x_2,y,i}| X_1(t)-X_2(t) |^2 \leq C|x_1 - x_2|^2 e^{C t}. 
\]
This implies, in view of Cauchy-Schwartz inequality, that
\beq{gronwall}
E^{s,x_1,x_2,y,i}| X_1(t)-X_2(t) | \leq C|x_1 - x_2| e^{Ct}. 
\eeq
Using this inequality and the Lipschitz continuity of $L$ and $\Phi$ with respect to the argument $x$, we have
\beq{PPRE}
 & & V(s,x_1,y,i)-V(s,x_2,y,i)\nonumber\\
&\leq&\sup_{u \in {\cal U}, \tau \in \Lambda }E^{s,x_1,x_2,y,i}\bigg[\int_s^\tau \bigg{|}L(t,X_1(t),Y(t),u(t),\al(t))-L(t,X_2(t),Y(t),u(t),\al(t))\bigg{|}dt\nonumber\\
 && \hspace{1 in}+|\Phi(\tau, X_1(\tau),Y(\tau),\al(\tau))-\Phi(\tau, X_2(\tau),Y(\tau),\al(\tau))|\bigg{]}\nonumber\\ 
&\leq&K \sup_{ \tau \in \Lambda }E^{s,x_1,x_2,y,i}\bigg[\int_s^\tau |X_1(t)-X_2(t)|dt+|X_1(\tau)-X_2(\tau)| \quad\hbox{for some }K>0 \nonumber \\
&\leq& KC|x_1 - x_2| \quad\hbox{for some constants }K>0, C>0. 
\eeq
The inequality (\ref{PPRE}) implies the (uniform) continuity of $V(s,x,y,i)$ with respect to $x$.
Next, we show the continuity of $V(s,x,y,i)$ with respect to $y$.\\
 Let $(s,x,y_1,i)$ and $(s,x,y_2, i)$ be two initial data such  that we have two solutions $(X,Y_1)$ and $(X,Y_2)$ with $Y_1(s)=y_1$, $Y_2(s)=y_2$ and $X=x$. Without loss of generality, we set ${\cal U }\equiv {\cal U}(s,x,y_1,i)\cap{\cal U}(s,x,y_2,i)$.  We will use the notation $E^{s,x,y_1,y_2,i}[\,\, \cdot\,\,]$ to represent $E[\,\, \cdot\,\,|X(s)=x,Y_1(s)=y_1,Y_2(s) = y_2, \al(s)=i]$.  For any control $u \in  {\cal U}$ and $t\in [s,T]$, using the Gronwall's inequality and the Lipschitz continuity of $L$ and $\Phi$ with respect to the arguments  $y$, we have
\beq{PPREII}
 & & V(s,x,y_1,i)-V(s,x,y_2,i)\nonumber\\
&\leq&\sup_{u \in {\cal U}, \tau \in \Lambda }E^{s,x,y_1,y_2,i}\bigg[\int_s^\tau \bigg{|}L(t,X(t),Y_1(t),u(t),\al(t))-L(t,X(t),Y_2(t),u(t),\al(t))\bigg{|}dt\nonumber\\
 && \hspace{1 in}+|\Phi(\tau, X(\tau),Y_1(\tau),\al(\tau))-\Phi(\tau, X(\tau),Y_2(\tau),\al(\tau))|\bigg{]}\nonumber\\ 
&\leq&K \sup_{ \tau \in \Lambda }E^{s,x,y_1,y_2,i}\bigg[\int_s^\tau |Y_1(t)-Y_2(t)|dt+|Y_1(\tau)-Y_2(\tau)| \quad\hbox{for some }K>0 \nonumber \\
&\leq& KC|y_1 - y_2| \quad\hbox{for some constants }K>0, C>0. 
\eeq
The inequality (\ref{PPREII}) implies the (uniform) continuity of $V(s,x,y,i)$ with respect to $y$.\\
Finally, we show the continuity of $V(s,x,y, i)$ with respect to $s$. 
Let $(X_t,Y_t)$ be the solution of 
(\ref{defi3}) that starts at $t=s$ with $X(s)=x$, $Y(s)=y$  and $\al(s)=i$. 
Let $0\leq s\leq s'\leq T$, we have
\beq{scont}
\disp
& & J(s,x,y,i;u,\tau)-J(s',x,y,i;u,\tau)\nonumber \\
&\leq&E^{s,x,y,i}\bigg[\int_s^{s'}e^{-r(t-s)} L(t,X(t),Y(t),u(t),\al(t))dt\nonumber \\
&&+\int_{s'}^{T}e^{-r(t-s)} L(t,X(t),Y(t),u(t),\al(t))dt -\int_{s'}^{T}e^{-r(t-s')} L(t,X(t),Y(t),u(t),\al(t))dt \bigg{]} \nonumber \\
&\leq& E^{s,x,y,i}\bigg[\int_s^{s'}e^{-r(t-s)} L(t,X(t),Y(t),u(t),\al(t))dt\nonumber \\
&&+\int_{s'}^{T}e^{-r(t-s')} L(t,X(t),Y(t),u(t),\al(t))dt -\int_{s'}^{T}e^{-r(t-s')} L(t,X(t),Y(t),u(t),\al(t)) dt\bigg{]} \nonumber \\
&\leq& E^{s,x,y,i}\bigg[\int_s^{s'}L(t,X(t),u(t),\al(t))dt \bigg{]}\nonumber \\
&\leq& C(s'-s) \quad \hbox{for some }C>0.
\eeq
Thus, we have 
\beq{sconti}
\disp
V(s,x,y,i)-V(s',x,y,i)&\leq&\sup_{u\in{\cal U}, \tau \in \Lambda}|J(s,x,y,i;u,\tau)-J(s',x,y,i;u,\tau)| \nonumber \\
&\leq& C|s'-s|.
\eeq
This guarantees the continuity of $V$ with respect to $s$.   
The linear growth inequality follows from the Lipschitz continuity of the value function with respect to $x$ and $y$. In fact,  there exist $K, C>0$ such that
\bea \disp
|V(s,x,y,i)|\leq K|x| +|V(s,0,y,i)|,
\eea
and
\bea
\disp
|V(s,0,y,i)|\leq C|y|+|V(s,0,0,i)|.
\eea
Combining the last two inequalities gives,
\bea
\disp
|V(s,x,y,i)|\leq \max(K,C)(|x|+|y|+|V(s,0,0,i)|)\leq C'(|x|+|y|+1),
\eea
for some $C'>\max(K,C)$.
This completes the proof. 
\endproof
\begin{rem}
It is well known from the Dynamic Programming Principle that, for any ${\cal{G}}_t$-stopping time $\tau$ we have 
\beq{mamcont}
\disp
V(s,x,y,i)=\sup_{u\in{\cal U}, \tau \in \Lambda_{s,T}}E^{s,x,y,i}\bigg[\int_s^{\tau}e^{-r(t-s)} L(t,X(t),Y(t),u(t),\al(t))dt \nonumber \\+ e^{-r(\tau-s)}V(\tau,X(\tau),Y(\tau),\al(\tau))\bigg{]}.
\eeq
\end{rem}
\begin{thm}\label{visco1}
The value function $V(s,x,y,i)$ is the unique
viscosity solution of equation (\ref{sys}).
\end{thm}
\paragraph{Proof.}
We first prove that $V$ is a viscosity supersolution of (\ref{sys}). We will verify that $V$ satisfies the inequality (\ref{max}).
Let $i \in {\cal M}$, and $\psi \in \mathnormal{C}^{1,2,1}([0,T]\times\rr\times\rr^+)$
such that $V(s,x,y,i)-\psi(s,x,y)$ has local minimum at $(s_0,x_0,y_0)$ in a neighborhood $N(s_0,x_0,y_0)$. Without loss of generality we assume that $V(s_0,x_0,y_0,\alpha_0)-\psi(s_0,x_0,y_0)=0.$
We set $\al(s_0) = \al_0$ and  define a function 
\beq{varphi2}
\varphi(s,x,y,i)=\left \{ \begin{array}{ll}
                    \psi(s,x,y), &  \textrm{if} \qquad i=\alpha_0,\\
V(s,x,y,i),  & \textrm{if} \qquad i \not =\alpha_0. \end{array} \right. 
\eeq
Let $\gamma\geq s_0 $ be the first jump time of $\alpha(\cdot)$ from the
initial state $\alpha(s_0)=\al_0$, and let $\theta \in [s_0,\gamma]$ be such that 
$(t,X(t),Y(t))$ starts at $(s_0,x_0,y_0)$ and stays in $N(s_0,x_0,y_0) $ for $s_0\leq t \leq \theta $.
Moreover, $ \alpha(t)= \alpha_0 $, for $ s_0\leq t\leq \theta$.
Let $u(\cdot)$ be an admissible control such that $u(t)= u$ for $t\in[s_0, \theta]$. 
From the Dynamical Programming Principle (\ref{mamcont}) we derive
\beq{inesup}
V(s_0,x_0,y_0,\al_0)\geq E^{s_0,x_0,y_0,\al_0}\bigg[\int_{s_0}^{\theta}e^{-r(t-s_0)} L(t,X(t),Y(t),u(t),\al(t))dt  \nonumber \\
+ e^{-r(\theta-s_0)}V(\theta,X(\theta),Y(\theta),\al(\theta))\bigg{]}.
\eeq
Using Dynkin's formula we have,
\beq{dynkin2}
& & E^{s_0,x_0,y_0,\al_0}[ e^{-r(\theta-s_0)} \varphi(\theta,X(\theta),Y(\theta),\alpha_0)]- \varphi(s_0,x_0,y_0,\alpha_0)  \nonumber\\
&=&E^{s_0,x_0,y_0,\al_0} \int_{s_0}^\theta e^{-r(t-s_0)}[-r\varphi(t,X(t),Y(t),\al_0)+({\cal L}^u\varphi)(t,X(t),Y(t),\al_0)]dt.
\eeq
where $({\cal L}^u\varphi)$ is defined as in (\ref{genra}). Note that $({\cal L}^u\varphi)(s,x,y,i)={\cal A}^{i,u}(\psi)(s,x,y)+ Q\varphi(s,x,y,\cdot)(i)$, where ${\cal A}^{i,u}(\psi)$ is given by
\bea
\disp
{\cal A}^{i,u}(\psi)(s,x,y)=\frac{\partial \psi(s,x,y)}{\partial s}+\frac{1}{2}\sigma^2(s,x,u,i)\frac{\partial^2\psi(s,x,y,i)}{\partial x^2}  \nonumber\\
\disp+  \mu(s,x,u,i)\frac{\partial \psi(s,x,y,i)}{\partial x}
 +\int_{\rr}\bigg(\psi(s,x+\gamma(s,x,u,i,z),y,i)\nonumber\\
\disp-\psi(s,x,y,i) -{\bf 1}_{\{|z|<1\}}(z)\frac{\partial \psi(s,x,y,i)}{\partial x}\cdot \gamma(s,x,u,i,z)\bigg)\nu(dz) 
\disp  \nonumber\\
\disp -  u\frac{\partial \psi(s,x,y,i)}{\partial y}.
\eea
Given that $(s_0,x_0,y_0)$ is the minimum of $V(t,x,y,\alpha_0)-\psi(t,x,y)$ in $N(s_0,x_0,y_0)$.
For $ s_0\leq t \leq \theta$, we have
\beq{vbig2}
 V(t,X(t),Y(t),\alpha_0) -\psi(t,X(t),Y(t)) &\geq& V(s_0,x_0,y_0,\alpha_0) -\psi(s_0,x_0,y_0)=0 \nonumber \\
  V(t,X(t),Y(t),\alpha_0) &\geq& \psi(t,X(t),Y(t))=\varphi(t,X(t),Y(t),\alpha_0).
\eeq 
Using equation (\ref{dynkin2}) and (\ref{vbig2}), we have  
\beq{dynkin3}
&  & E^{s_0,x_0,y_0,\al_0}[ e^{-r(\theta-s_0)}V(\theta,X(\theta),Y(\theta),\alpha_0)]- V(s_0,x_0,y_0,\alpha_0)\nonumber \\
&\geq& E^{s_0,x_0,y_0,\al_0} \int_{s_0}^\theta e^{-r(t-s_0)} \bigg[ {\cal A}^{\al_{s_0},u}(\psi)(t,X(t),Y(t))+Q\varphi(t,X(t),Y(t),\cdot)(\al_0)\nonumber \\
& &-rV(t,X(t),Y(t),\al_0)\bigg]dt. 
\eeq
Moreover, we have 
\beq{Qarray2}
Q\varphi(t,X(t),Y(t),\cdot)(\alpha_0) &=& \sum_{\beta \not = \alpha_0}q_{\alpha_0 \beta}\Bigg{(}\varphi(t,X(t),Y(t),\beta)-\varphi(t,X(t),Y(t),\alpha_0)\bigg{)}\nonumber \\
&\geq& \sum_{\beta \not = \alpha_0}q_{\al_0 \beta}\Bigg{(}V(t,X(t),Y(t),\beta)-V(t,X(t),Y(t),\alpha_0)\bigg{)}\nonumber\\
&\geq&QV(t,X(t),Y(t),\cdot)(\alpha_0).
\eeq
Combining (\ref{dynkin3}) and (\ref{Qarray2}), we have
 \beq{eeer}
& &E^{s_0,x_0,y_0,\al_0} e^{-r(\theta-s_0)}[V(\theta,X(\theta),Y(\theta),\alpha_0)]- V(s_0,x_0,y_0,\alpha_0)\nonumber \\
&\geq& E^{s_0,x_0,y_0,\al_0} \int_{s_0}^\theta e^{-r(t-s_0)} \bigg[ {\cal A}^{\al_{s_0},u}(\psi)(t,X(t),Y(t))\nonumber \\&&\hspace{0.5 in}+QV(s_0,x_0,y_0\cdot)(\al_s)-rV(t,X(t),Y(t),\al_s)\bigg]dt. 
\eeq
It follows from (\ref{inesup}) and (\ref{eeer}) that
\bea
\disp
E^{s_0,x_0,y_0,\al_0}\int_{s_0}^\theta\frac{ e^{-r(t-s_0)}}{\theta-s_0} \bigg[ {\cal A}^{\al_{0},u}(\psi)(t,X(t),Y(t))+QV(t,X(t),Y(t),\cdot)(\al_0)\\-rV(t,X(t),Y(t),\al_0)+L(t,X(t),Y(t),u(t),\al(t))\bigg]dt \leq 0.
\eea
Sending  $\theta \rightarrow s_0$ leads to
\beq{super-ineq}
-rV(s_0,x_0,y_0,\al_0)+{\cal A}^{\al_{0},u}(\psi)(s_0,x_0,y_0)\nonumber \\ 
+QV(s_0,x_0,y_0,\cdot)(\al_0)+L(s_0,x_0,y_0,u,\al_0)\leq 0.
\eeq
Since this inequality is true for any arbitrary control $u(t)\equiv u$ in ${\cal U}(s,x_s,\al_s)$, then taking the supremum over all values $u\in U$ we have
\beq{Supersol1}
\disp
rV(s_0,x_0,y_0,\al_0) -\sup_{u\in{\cal U}}\Bigg({\cal A}^{\al_0,u}(\psi)(s_0,x_0,y_0)\nonumber \\
+QV(s_0,x_0,y_0,\cdot)(\al_0)+L(s_0,x_0,y_0,u,\al_0)\Bigg)\geq 0.
\eeq
Moreover it follows from the definition of the value function (\ref{val2}) that for any control $u \in {\cal U}$,
\beq{Supersol2}
\disp V(s_0,x_0,y_0,\al_0) \geq J(s_0,x_0,y_0,\al_0;u,s_0) = \Phi(s_0,x_0,y_0,\al_0).
\eeq
Combining the inequalities (\ref{Supersol1}) and (\ref{Supersol2}) obviously proves  that the value function $V$ is a viscosity  supersolution  as defined in (\ref{min}).\\
Now, let us prove the subsolution inequality (\ref{min}). First,  we want to show that for each $\iota \in {\cal M}$ ,
\beq{rrmax}
rV(s_0,x_0,y_0,\iota) -\sup_{u\in{\cal U}}\Bigg({\cal A}^{\al_0,u}(\psi)(s_0,x_0,y_0)\nonumber \\
+QV(s_0,x_0,y_0,\cdot)(\iota)+L(s_0,x_0,y_0,u,\iota)\Bigg) \leq 0,
\eeq 
where $(s_0,x_0,y_0)$ is a local maximum of $V(s,x,y,\iota)-\psi(s,x,y)$ .
Let us assume otherwise that the inequality (\ref{rrmax}) does not hold. In other terms, we assume that we can find a state $\al_0 \in {\cal M}$, values $(s_0,x_0,y_0)$ and a function $\phi\in {\cal C}^{1,2}([0,T]\times \rr\times\rr^+)$ such that $V(t,x,y,\alpha_0)-\phi(t,x,y)$ has a local maximum at $ (s_0,x_0,y_0)\in [0,T]\times \rr\times\rr^+,$  and we have
\beq{absurd}
rV(s_0,x_0,y_0,\al_0) -\sup_{u\in{\cal U}}\Bigg({\cal A}^{\al_0,u}(\psi)(s_0,x_0,y_0)\nonumber \\
+QV(s_0,x_0,y_0,\cdot)(\al_0)+L(s_0,x_0,y_0,u,\al_0)\Bigg)\geq \delta.
\eeq
for some constant $\delta>0$.\\
Let us assume without loss of generality that $V(s_0,x_0,y_0,\alpha_0)-\phi(s_0,x_0,y_0)= 0$.
We define
\beq{varpsi10}
\Psi(s,x,y,i)=\left \{ \begin{array}{ll}
                    \phi(s,x,y), &  \textrm{if} \qquad i=\alpha_0,\\
V(s,x,y,i),  & \textrm{if} \qquad i \not =\alpha_0 .\end{array} \right. 
\eeq
Let $\gamma $ be the first jump time of $\alpha(\cdot)$ from the state $\alpha_0$, and let $\theta_0 \in [s_0,\gamma]$ be such that $(t,X(t),Y(t))$ starts at $(s_0,x_0,y_0)$ and stays in $N(s_0,x_0,y_0) $ for $s_0\leq t \leq \theta_0 $.
Since $\theta_0 \leq \gamma $ we have $ \alpha(t)= \alpha_0 $, for $ s_0\leq t\leq \theta_0$.
Moreover, since $V(s_0,x_0,y_0,\alpha_0)-\phi(s_0,x_0,y_0)= 0$ and attains its maximum at $(s_0,x_0,y_0)$ in $N(s_0,x_0,y_0)$ then 
\[
V(\theta,X(\theta),Y(\theta),\alpha(\theta))\leq \phi(\theta,X(\theta),Y(\theta))\quad \hbox{for  }\,\,\, s_0\leq \theta \leq \theta_0.
\]  
 Thus, we also have
 \beq{Phi}
V(\theta,X(\theta),Y(\theta),\alpha(\theta))\leq \Psi(\theta,X(\theta),Y(\theta),\alpha(\theta))\quad \hbox{for  }\,\,\, s_0\leq \theta \leq \theta_0.
\eeq
Using the Dynamical Programming Principle (\ref{mamcont}), it clear that for any admissible control $u(\cdot)$ and stopping time $\tau$ such that $s_0<\tau \leq \theta_0$, we have 
\bea \disp
 J(s_0,x_0,y_0,\al_{0};u,\tau)&\leq \disp E^{s_0,x_0,y_0,\al_0}\bigg[\int_{s_0}^{\tau}e^{-r(t-s_0)} L(t,X(t),Y(t),u(t),\al(t))dt \\
&  + e^{-r(\tau-s_0)}V(\tau,X(\tau),Y(\tau),\al(\tau))\bigg{]}\\
 & \disp \leq E^{s_0,x_0,y_0,\al_0}\bigg[\int_{s_0}^{\tau}e^{-r(t-s_0)} L(t,X(t),u(t),\al(t))dt\\
& + e^{-r(\tau-s_0)}\Psi(\tau,X(\tau),Y(\tau),\al(\tau))\bigg{]}.
 \eea
 Note that
\beq{Qmat21}
Q\Psi(t,X(t),Y(t),\cdot)(\alpha_0) & =  & \sum_{\beta \not = \alpha_0}q_{ \alpha_0 \beta}(V(t,X(t),Y(t),\beta)-\phi(t,X(t),Y(t))) \nonumber \\
& \leq  & \sum_{\beta \not = \alpha_0}q_{ \alpha_0 \beta}(V(t,X(t),Y(t),\beta)- V(t,X(t),Y(t),\alpha_0))  \nonumber \\
& \leq & QV(t,X(t),Y(t),\cdot)(\alpha_0).
\eeq
 Using the inequality (\ref{absurd}) we have 
 \beq{new-ineq}
& &J(s_0,x_0,y_0,\al_0;u,\tau) \nonumber \\
&\leq& \disp E^{s_0,x_0,y_0,\al_0}\bigg(\int_{s_0}^{\tau}e^{-r(t-s_0)}\bigg{\{} -\delta +rV(t,X(t),Y(t),\al_0)-{\cal A}^{\al_0,u}(\phi)(t,X(t),Y(t))
\nonumber\\
&&-  QV(t,X(t),Y(t),\cdot)(\al_0) \bigg{\}}dt\disp+e^{-r(\tau-s_0)}\Psi(\tau,X(\tau),Y(\tau),\al_0)\bigg{)}.
\eeq
 The Dynkin's formula, (\ref{varpsi10}) and (\ref{Qmat21})   imply that  
\beq{ligne20}
& & E^{s_0,x_0,y_0,\al_0} e^{-r(\tau-s_0)}\Psi(\tau,X(\tau),Y(\tau),\alpha_0) \nonumber\\
&=& E^{s_0,x_0,y_0,\al_0} \int_{s_0}^\tau e^{-r(t-s_0)} \Bigg{[}{\cal A}^{\al_0,u}(\phi)(t,X(t),Y(t))+ Q\Psi(t,X(t),Y(t),\cdot)(\alpha_0) \nonumber \\
& &-r\Psi(t,X(t),Y(t),\alpha_0) \bigg{]}dt  +\Psi(s_0,x_0,y_0,\alpha_0) \nonumber\\
&\leq& E^{s_0,x_0,y_0,\al_0} \int_{s_0}^\tau e^{-r(t-s_0)} \Bigg{[}{\cal A}^{\al_0,u}(\phi)(t,X(t),Y(t))+ QV(t,X(t),Y(t),\cdot)(\alpha_0) \nonumber\\
 &&-rV(t,X(t),Y(t),\alpha_0)  \bigg{]}dt+V(s_0,x_0,y_0,\alpha_0).
\eeq
Combining (\ref{new-ineq}) and (\ref{ligne20}) we have
\beq{end}
J(s_0,x_0,y_0,\al_0;u,\tau)\leq  E^{s_0,x_0,y_0,\al_0}\bigg(-\int_{s_0}^{\tau}e^{-r(t-s_0)}\delta dt\bigg)+V(s_0,x_0,y_0,\alpha_0).
\eeq
It is easy to see that the quantity $\gamma=\disp E^{s_0,x_0,y_0,\al_0}\bigg(\int_{s_0}^{\tau}e^{-r(t-s_0)}\delta dt\bigg)>0$, thus taking the supremum over all admissible control $u(\cdot)\equiv u$  and stopping time $\tau \in \Lambda$ on (\ref{end}) we obtain
\beq{finish}
V(s_0,x_0,y_0,\al_0)\leq -\gamma +V(s_0,x_0,y_0,\al_0),
\eeq
which is a contradiction. This proves that the inequality (\ref{rrmax}) is satisfied. Obviously, we derive the subsolution inequality (\ref{min}). 
Therefore, $V$ is a viscosity solution of (\ref{sys}). The uniqueness of the viscosity solution can be obtained as in  Biswas et al. (2010)  by applying nonlocal extensions of the Jensen-Ishii Lemma. For more on the derivation of the maximum principle for nonlocal operators, one can also refer to Barles and Imbert (2008).
\endproof

\subsection{Optimal extraction and stopping strategies}
In this section, we prove a verification theorem for this particular class of control problems. Even though this result is quite standard in control theory, we provide the proof of a verification theorem here because we did not find a reference with a similar result. Moreover, this result will give us the basis for numerically identifying optimal extraction and stopping policies. Let ${\cal P}^{1,2,1,+}(W(s,x,y,i))$ and ${\cal P}^{1,2,1,-}(W(s,x,y,i))$ be respectively the parabolic superjet and subjet of $W(s,x,y,i)$ respectively and ${\cal \bar{P}}^{1,2,1+}(W(s,x,y,i))$,\\  ${\cal \bar{P}}^{1,2,1-}(W(s,x,y,i))$ their respective closure.
\begin{thm}\label{Verification}
Let $W$ be such that for  each $ i \in {\cal M}$,   $W(\cdot,\cdot,\cdot,i)\in C([0,T],\rr,\rr^+)$  and satisfies  the assumptions (\ref{condition1}) and (\ref{condition2}), and assume that $W$ is the viscosity solution of the   HJB equation (\ref{sys}).  Moreover let  $D=\{(t,x,y,i)\in [0,T]\times\rr\times\rr^+\times{\cal M}, \quad \Phi(t,x,y,i) < W(t,x,y,i)\}$. The set $D$ is called the continuation region. Then,\\
\begin{enumerate}
\item  for all $u \in {\cal U}(s,x,y,i)$ and stopping time  $\tau \in \Delta_{s,T}$ we have 
\beq{verification_1}
 W(s,x,y,i)\geq J(s,x,y,i;u,\tau).
 \eeq
\item If there exists an admissible control $u^*(\cdot) \in {\cal U}(s,x,y,i)$ and paths $(X^*(\cdot), Y^*(\cdot))$ such that \\
$(p(t),q(t),\rho(t),M(t)) \in {\cal \bar{P}}^{1,2,1+}(W(t,X^*(t),Y^*(t),\al(t))\cup{\cal \bar{P}}^{1,2,1-}(W(t,X^*(t),Y^*(t),i) $, a.e. $t\in [s,T]$ and $(t, X^*(t), Y^*(t),i) \in D$ 
\beq{differential_2}
u^*(t) \in \arg\max\bigg[
p(t)+\frac{1}{2}\sigma^2(t,X^*(t),u,i)M(t) +  \mu(t,X^*(t),u,i)q(t) \\
 -  u\rho(t) +\int_{\rr}\bigg(W(t,X^*(t)+\gamma(t,X^*(t),u,i,z),Y^*(t),\al(t))\nonumber\\
-W(t,X^*(t),Y^*(t),i) -{\bf 1}_{\{|z|<1\}}(z)p(t)\cdot \gamma(t,X^*(t),u,i,z)\bigg)\nu(dz) 
\disp  \nonumber\\ +  QW(t,X^*(t),Y^*(t),\cdot)(i) + L(t, X^*(t), Y^*(t), u(t), i)\nonumber
 \bigg],
\eeq
for a.e.  $t\in [s,T]$ 
then $u^*$ is an optimal control and $\tau_D = \inf\{t>0, (t, X^*(t), Y^*(t), \al(t)) \not \in D\} $ is the optimal stopping time.   
\end{enumerate}
\end{thm}
Before proving this theorem we state the following lemma, its proof can be found in Fleming and  Soner (2006).
\begin{lem}\label{lemmo}
Let $f(\cdot,\cdot,\cdot,i)\in C([0,T]\times\rr\times\rr^+)$ for each $i \in  {\cal M}$, ${\cal P}^{1,2,1,+}(W(s,x,y,i)$ \\
$(resp.\,\,\, {\cal P}^{1,2,1,-}(W(s,x,y,i))$  consists  of the set of $(\frac{\partial \phi(s,x,y)}{\partial s}, \frac{\partial \phi(s,x,y)}{\partial x}, \frac{\partial \phi(s,x,y)}{\partial y}, \frac{\partial^2 \phi(s,x,y)}{\partial x^2})$ where \\$\phi\in C^{1,2,1}([0,T]\times \rr \times \rr^+)$ and  $f-\phi$ has a
global maximum (resp. minimum) at $ (s,x,y)$.   
\end{lem}   
\paragraph{Proof.}
First of all, (\ref{verification_1}) follows from the uniqueness of the HJB equation (\ref{sys}).  Let $ u^*$ be an admissible control $u^*(\cdot) \in {\cal U}(s,x,y,i)$ and  $(X^*(\cdot), Y^*(\cdot))$ be a solution of (\ref{defi3}) such that 
$(t, X^*(t), Y^*(t),\al(t)) \in D$ and $(p(t),q(t),\rho(t),M(t)) \in {\cal \bar{P}}^{1,2,+}(W(t,X^*(t),Y^*(t),\al(t))$, for almost every $t\in [s,T]$. Using \lemref{lemmo} we know that there exists a sequence of  a smooth functions $\phi_n\in C^{1,2,1}([0,T]\times \rr \times \rr^+)$ such that  $W-\phi_n$ has a
global minimum  at $ (s,X^*(s),Y^*(s))$  and that 
\bea
\disp  &(p(t),q(t),\rho(t),M(t)) \\
&\disp = \lim_{n\rightarrow \infty} (\frac{\partial \phi_n(t,X^*(t),Y^*(t))}{\partial s}, \frac{\partial \phi_n(t,X^*(t),Y^*(t))}{\partial x}, \frac{\partial \phi_n(t,X^*(t),Y^*(t))}{\partial y}, \frac{\partial^2 \phi_n(t,X^*(t),Y^*(t))}{\partial x^2}).
\eea
Without loss of generality we can assume that for each $n$
\[
W(s,X^*(s),Y^*(s),\al(s))- \phi_n(s,X^*(s),Y^*(s)))=0
\]
and define the function $\varphi_n$ as follows 
\bea
\varphi_n(s,x,y,i)=\left \{ \begin{array}{ll}
                    \phi_n(s,x,y), &  \textrm{if} \qquad i=\alpha(s),\\
W(s,x,y,i),  & \textrm{if} \qquad i \not =\alpha(s) .\end{array} \right. 
\eea
 Therefore,  for $t \in [s,T]$  we have 
\beq{lastdynk1}
& W(t,X^*(t),Y^*(t),\al(t))- \varphi_n(t,X^*(t),Y^*(t)),\al(t))\\
&\geq W(s,X^*(s),Y^*(s),\al(s))- \phi_n(s,X^*(s),Y^*(s)))=0. \nonumber
\eeq
It comes from (\ref{lastdynk1}) that 
\beq{lastdynk2}
&&Q\varphi_n(t,X^*(t),Y^*(t),\cdot)(\al(s)) \nonumber \\
&=&  \sum_{\beta\not = \al(s)}q_{\al(s)\beta}(\varphi_n(t,X^*(t),Y^*(t),\beta) - \varphi_n(t,X^*(t),Y^*(t),\al(s) \nonumber\\
                            &\geq&   \sum_{\beta\not = \al(s)}q_{\al(s)\beta}(W(t,X^*(t),Y^*(t),\beta) - W(t,X^*(t),Y^*(t),\al(s))\nonumber\\
&=& QW(t,X^*(t),Y^*(t),\cdot)(\al(s))
\eeq
In addition, it is obvious that $\Phi(\tau_D, X^*(\tau_D),Y^*(\tau_D), \al(\tau_D))=W(\tau_D, X^*(\tau_D),Y^*(\tau_D), \al(\tau_D))$, thus using   the Dynkin formula we have
\bea
 & E[ e^{-r(\tau_D-s)}\varphi_n(\tau_D,X^*(\tau_D),Y^*(\tau_D),\al(\tau_D))- \varphi_n(s,X^*(s),Y^*(s),\al(s))]\nonumber \\
&\disp = E\int_{s}^{\tau_D} e^{-r(\xi-s)} \bigg[ {\cal A}^{\al(s),u^*(\xi)}(\phi_n)(\xi,X^*(\xi),Y^*(\xi))+Q\varphi_n(\xi,X^*(\xi),Y^*(\xi),\cdot)(\al(s))\nonumber \\
& -r\varphi_n(\xi,X^*(\xi),Y^*(\xi),\al(s))\bigg]d\xi. 
\eea
So using the fact that $W(s,X^*(s),Y^*(s),\al(s))=\varphi_n(s,X^*(s),Y^*(s),\al(s))$  and the inequalities (\ref{lastdynk1}), (\ref{lastdynk2})  we have
\beq{lastdicth}
&&W(s,X^*(s),Y^*(s),\al(s))\nonumber \\
& =& E\int_{s}^{\tau_D} e^{-r(\xi-s)} \bigg[ -{\cal A}^{\al(s),u^*(\xi)}(\phi_n)(\xi,X^*(\xi),Y^*(\xi))-Q(u^*)\varphi_n(\xi,X^*(\xi),Y^*(\xi),\cdot)(\al(s))\nonumber \\
& &+r\varphi_n(\xi,X^*(\xi),Y^*(\xi),\al(s))\bigg]d\xi +  E[ e^{-r(\tau_D-t)}\varphi_n(\tau_D,X^*(\tau_D),Y^*(\tau_D),\al(\tau_D))]\nonumber\\
&\leq& E\int_{s}^{\tau_D} e^{-r(\xi-s)} \bigg[ -{\cal A}^{\al(s),u}(\phi_n)(\xi,X^*(\xi),Y^*(\xi))-QW(\xi,X^*(\xi),Y^*(\xi),\cdot)(\al(s))\nonumber \\
& & +rW(\xi,X^*(\xi),Y^*(\xi),\al(s))\bigg]d\xi +  E[ e^{-r(\tau_D-s)}\Phi(\tau_D,X^*(\tau_D),Y^*(\tau_D),\al(\tau_D))].
\eeq
Taking the limit as $n$ goes to infinity in the last inequality we have
\bea
&W(s,X^*(s),Y^*(s),\al(s))\\
&\disp\leq  E\int_{s}^{\tau_D} e^{-r(\xi-s)} \Bigg(-\bigg[
p(\xi)+\frac{1}{2}\sigma^2(\xi,X^*(\xi),u^*(\xi),\al(\xi))M(\xi)\\
&\disp +  \mu(\xi,X^*(t),u^*(\xi),\al(s))q(\xi)  - u^*(\xi)\rho(\xi) +\int_{\rr}\big(W(\xi,X^*(\xi)\nonumber\\
&\disp+\gamma(\xi,X^*(\xi),u^*(\xi),\al(s),z),Y^*(\xi),\al(s))-W(\xi,X^*(\xi),Y^*(\xi),\al(s))  \\
&\disp-{\bf 1}_{\{|z|<1\}}(z)p(\xi)\cdot \gamma(\xi,X^*(\xi),u^*(\xi),\al(s),z)\big)\nu(dz)\nonumber\\
&  \disp  + L(\xi, X^*(\xi), Y^*(\xi), u(\xi), \al(s))  +QW(\xi,X^*(\xi),Y^*(\xi),\cdot)(\al(s))\bigg] \nonumber\\
&  \disp +rW(\xi,X^*(\xi),Y^*(\xi),\al(s))\Bigg)d\xi +  E[ e^{-r(\tau_D-s)}\Phi(\tau_D,X^*(\tau_D),Y^*(\tau_D),\al(\tau_D)) \bigg].
\eea
Taking into account (\ref{differential_2}) we obtain
\bea
&W(s,X^*(s),Y^*(s),\al(s))\\
&\disp\leq  E\bigg[\int_{s}^{\tau_D} e^{-r(\xi-s)}L(\xi,X^*(\xi),Y^*(\xi),u^*(\xi),\al(s))d\xi\nonumber\\
&  \disp \qquad+   e^{-r(\tau_D-s)}\Phi(\tau_D,X^*(\tau_D),Y^*(\tau_D),\al(\tau_D)) \bigg]\\
&\disp= J(s,X^*(s),Y^*(s),\al(s);u^*,\tau_D).
\eea
Therefore $u^*$ is the optimal control and $\tau_D$ is the optimal stopping time. This ends the proof.  
\endproof
\section{Numerical Approximation}
In this section, we construct an explicit finite difference scheme and
show that it converges to the unique viscosity solution of equation (\ref{sys}).  Let $k, h, l\in(0,1) $ be the step size with respect to $s$, $x$ and $y$ respectively,  we consider the following  finite difference
operators  $\Delta_s$, $\Delta_x$, $\Delta_{xx}$ and $\Delta_y$ defined by
\bea \ad \Delta_s V(s,x,y,i)=\frac{V(s+k,x,y,i)-V(s,x,y,i)}{k}, \quad \Delta_x V(s,x,y,i)=\frac{V(s,x+h,y,i)-V(s,x,y,i)}{h} \\
\ad \Delta_y V(s,x,y,i) =\frac{V(s,x,y+l,i)-V(s,x,y,i)}{l},\\ 
\ad \Delta_{xx} V(s,x,y,i)=\frac{V(s,x+h,y,i)+ V(s,x-h,y,i)-2V(s,x,y,i)}{h^2}.
 \eea
 Note that for each $u \in U$, the generator ${\cal L}^u$ defined (\ref{genra}) of the Markov process $(X(t),Y(t),\al(t)$ can be rewritten as follows 
\bea
{\cal L}^uf(s,x,y,i) = {\rm D}f(s,x,y,i;u) + {\rm I}f(s,x,y,i;u)  + Qf(s,x,y,\cdot)(i), 
\eea  
where ${\rm D}f$ is the differential part and ${\rm I}f$ is the integral part. We will approximate ${\rm I}f$ using the Simpson quadrature. In fact we have 
\bea
\ad{\rm I}^uf(s,x,y,i)\\
\ad = \int_{\rr}\bigg(f(s,x+\gamma(s,x,u,i,z),y,i)-f(s,x,y,i) -{\bf 1}_{\{|z|<1\}}(z)\frac{\partial f(s,x,y,i)}{\partial x}\cdot \gamma(s,x,u,i,z)\bigg)\nu(dz). 
\eea
Using (\ref{finiteIntensity}), we know that  $\disp \Gamma = \int_{\rr} \nu(dz)<\infty$. Therefore the integral part of the generator can be simplified as follows;
\bea
\ad{\rm I}^uf(s,x,y,i)\\
\ad = \int_{\rr}f(s,x+\gamma(s,x,u,i,z),y,i)\nu(dz)-\frac{\partial f(s,x,y,i)}{\partial x}\int_{-1}^1 \gamma(s,x,u,i,z)\nu(dz)  -f(s,x,y,i)\Gamma. 
\eea
Let $\xi\in(0,1)$ the step size of the Simpson's quadrature, the corresponding approximation of the the integral part is
\bea
\ad{\rm I}^u_{\xi}f(s,x,y,i)\\
\ad = \sum_{j=0}^{N_\xi}c_jf(s,x+\gamma(s,x,u,i,z_j),y,i)-\frac{\partial f(s,x,y,i)}{\partial x}\sum_{j=0}^{M_\xi} d_j\gamma(s,x,u,i,z_j)-f(s,x,y,i)\Gamma, 
\eea
where the $(c_j)_{0\leq j\leq N_\xi}$ and $(d_j)_{0\leq j\leq M_\xi}$ are the corresponding sequences of the coefficients of the Simpson's quadrature. In fact $\disp  \lim_{N_\xi \rightarrow \infty}  \sum_{j=0}^{N_\xi}c_j =\Gamma$ and $\disp \lim_{M_\xi \rightarrow \infty}\sum_{j=0}^{M_\xi}d_j=\int_{-1}^1\nu(dz) $.
The corresponding discrete version of equation (\ref{sys}) is given by
\beq{disc-HJB}
 V(s,x,y,i) &=&\max\bigg[\frac{1}{r}\bigg(\Delta_s V(s,x,y,i)+\sup_{u\in U} \Bigg(\frac{1}{2}\sigma^2(s,x,u,i)\Delta_{xx}V(s,x,y,i)\\
&&+  \mu(s,x,u,i)\Delta_x V(s,x,y,i) +{\rm I}^u_{\xi}V(s,x,y,i)
  + QV(s,x,y,\cdot)(i)\Bigg)  \nonumber \\
&& - u\Delta_yV(s,x,y,i) \bigg), \Phi(s,x,y,i)  \Bigg{]}. \nonumber
\eeq
First we prove the  existence of a solution for the discretized equation on bounded subsets of the domain of study $[0,T]\times {\cal D}$ where  ${\cal  D}:=\rr \times \rr^+ \times {\cal M}$. Taking into account 
the fact that we can reasonably assume that the both the commodity price, as well as the size of the remaining reserve of the mineral, cannot go beyond a certain threshold. We can 
without loss of generality study this discretized problem on truncated subsets ${\cal D}_K$ of the domain ${\cal D}$ such that  ${\cal D}_K :=\{ (x,y,i) \in {\cal D}, x\leq K, y\leq K\}$.   

\begin{lem}\label{DiscreteLemma}
There exists a  unique bounded function $V_{l,h,k}$  defined on $[0,T]\times {\cal D}_K$  that solves equation (\ref{disc-HJB}) with terminal condition $V_{l,h,k}(T,x,y,i) = \Phi(T,x,y,i)$, for each $\xi>0$ small enough.
\end{lem}
\paragraph{Proof.}
We define the operator ${\cal F}_\xi$  on bounded functions on $[0,T]\times {\cal D}_K$ as follows
\beq{DiscrOpe}
&&{\cal F}_\xi(V)( s,x,y,i; h,k,l)\nonumber \\
 &=& \max\bigg[ \frac{1}{rk} V(s+k,x,y,i)+\sup_{u\in U}\Bigg( a(s,x,i;u)V(s,x+h,y,i) \nonumber \\
&&+ b(s,x,i;u)V(s,x-h,y,i) -c(s,x,y,i;u)V(s,x,y,i)  - \frac{u}{rl}V(s,x,y+l,i) \nonumber\\
&&
+ \sum_{j=0}^{N_\xi}\frac{c_j}{r}V(s,x+\gamma(s,x,u,i,z_j),y,i)+ \sum_{j\ne i} \frac{q_{ij}}{r}V(s,x,y,j)\nonumber\\
&&- V(s,x+h,y,i)\frac{ \sum_{j=0}^{M_\xi} d_j\gamma(s,x,u,i,z_j)}{rh}\Bigg), \Phi(s,x,y,i) \bigg], \\
&&\hbox{if }(s,x,y,i)\in [0,T)\times {\cal D}_K, \nonumber\\
&&{\cal F}_\xi(V)(T,x,y,i;h,k,l) = \Phi(T,x,y,i).\nonumber
\eeq 
Where the coefficients $a(s,x,i;u), b(s,x,i;u)$ and $c(s,x,y,i;u)$ are defined as follows 
\bea
\ad c(s,x,y,i;u) = \frac{1}{rk} + \frac{\sigma^2(s,x,u,i)}{rh^2} +  \frac{\mu(s,x,u,i)}{rh} -\frac{ \sum_{j=0}^{M_\xi} d_j\gamma(s,x,u,i,z_j)}{rh}- \frac{u}{rl} +\frac{\Gamma}{r} +\sum_{j\ne i} \frac{q_{ij}}{r}, \\  
\ad a(s,x,i;u) = \frac{\sigma^2(s,x,u,i)}{2 r h^2}  + \frac{\mu(s,x,u,i)}{r h},\\
\ad b(s,x,i;u)  =\frac{\sigma^2(s,x,u,i)}{2 r h^2}.
 \eea
Note that equation (\ref{disc-HJB}) is equivalent to $ V(s,x,y,i) = {\cal F}_\xi(V)(s,x,y,i;h,k,l)$, it suffices to show the operator ${\cal F}_\xi$ has a fixed point.
Using the fact that the difference of supremums is less than the supremum of differences, if we have two bounded functions $V, W$ on $ [0,T]\times {\cal D}_K$, it is clear that 
\bea
\ad |{\cal F}_\xi(V)(s,x,y,i;h,k,l) - {\cal F}_\xi(W)(s,x,y,i;h,k,l)| \\
\ad \leq  \bigg |  \sup_{u \in U} \bigg[  \bigg( a(s,x,i;u) + b(s,x,i;u) - c(s,x,y,i;u) + \frac{1}{rk}  +  \sum_{j=0}^{N_\xi}\frac{c_j}{r}  + \sum_{j\ne i} \frac{q_{ij}}{r}\\
\ad - \frac{u}{rl}-\frac{ \sum_{j=0}^{M_\xi} d_j\gamma(s,x,u,i,z_j)}{rh}\bigg)\sup_{[0,T]\times{\cal D}_T}|V-W|\bigg] \bigg|\\
\ad \leq \bigg|\sum_{j=0}^{N_\xi}\frac{c_j}{r}-\frac{\Gamma}{r}\bigg|\sup_{[0,T]\times{\cal D}_T}|V-W|.
\eea
Consequently, for $\xi \in(0,1) $  small enough so that $\disp \bigg|\sum_{j=0}^{N_\xi}\frac{c_j}{r}-\frac{\Gamma}{r}\bigg| <1$, the map ${\cal F}_\xi$ is a contraction on the space of bounded functions on $[0,T]\times {\cal D}_K$, using the Banach's fixed point theorem we conclude the proof of the lemma.
\endproof
\begin{rem}
\begin{enumerate}
\item Define  $S\to (0,1)^4\times[0,T]\times\rr\times\rr^+\times {\cal M}\times \rr\times B([0,T]\times\rr\times\rr^+\times {\cal M} )$ as follows;
\beq{schema}
&& S(\xi, h,k,l, s,x,y,i,w,W)\nonumber \\
&=& w - \max\bigg[ \frac{1}{rk} W(s+k,x,y,i)+\sup_{u\in U}\Bigg( a(s,x,i;u)W(s,x+h,y,i) \nonumber \\
&&+ b(s,x,i;u)W(s,x-h,y,i) -c(s,x,y,i;u)w  - \frac{u}{rl}W(s,x,y+l,i) \nonumber\\
&&
+ \sum_{j=0}^{N_\xi}\frac{c_j}{r}W(s,x+\gamma(s,x,u,i,z_j),y,i)+ \sum_{j\ne i} \frac{q_{ij}}{r}W(s,x,y,j)\nonumber\\
&&- W(s,x+h,y,i)\frac{ \sum_{j=0}^{M_\xi} d_j\gamma(s,x,u,i,z_j)}{rh}\Bigg), \Phi(s,x,y,i) \bigg].
\eeq
Obviously $V_{h,k,l}$ solves the equation $S(\xi, h,k,l, s,x,y,i,V_{h,k,l}(s,x,y,i),V_{h,k,l}) =0$.
 It is clear that for $h$ small enough the coefficients $a(s,x,i;u)>0$, $b(s,x,i;u)>0$ therefore the scheme $S$ is monotone with respect to argument $W$,  i.e., for  all $\xi,h,k,l \in(0,1),s\in [0,T], x\in \rr, y\in \rr^+, i \in {\cal M}$ and $W_1, W_2 \in  B([0,T]\times\rr\times\rr^+\times {\cal M} )$, we have
\beq{monotone}
\quad S(\xi, h,k,l, s,x,y,i,w,W_2)\leq S(\xi, h,k,l, s,x,y,i,w,W_1) \quad \hbox{whenever} \quad W_1\leq W_2.
\eeq
\item 
\lemref{DiscreteLemma} implies that the our finite difference scheme is stable since the solution of the scheme is bounded independently of the step sizes $h,k, l\in(0,1)$. Moreover is obvious to see that the scheme is consistent. We have the following convergence theorem.
\end{enumerate}

\end{rem}
\begin{thm}
For each $\xi>0$ small enough, let $V_{h,k,l}$ be the solution of the discrete scheme obtained in \lemref{DiscreteLemma}. Then as $\xi \downarrow 0$ and  $(h,k,l)\rightarrow \infty$ the sequence $V_{h,k,l}$ converges locally uniformly on $[0,T]\times {\cal D}$ to the unique viscosity solution $V$ of (\ref{sys}).
\end{thm}

\paragraph{Proof}
Define
\beq{sup} \barray
V^{*}(s,x,y,i)\ad=\limsup_{\theta \to s,\eta \to
x,\zeta\to y, k\downarrow 0, h \downarrow 0, l\downarrow 0}
V_{k,h, l}(\theta,\eta,\zeta,i) \,\, \hbox{and}  \\
V_{*}(s,x,y,i)\ad=\liminf_{\theta \to s,\eta \to
x,\zeta\to y, k\downarrow 0, h \downarrow 0, l\downarrow 0}
V_{k,h, l}(\theta,\eta,\zeta,i) \,\, .
\earray\eeq
We claim that $ V^{*}$ and $V_{*}$
are  sub- and supersolutions of (\ref{sys}), respectively.

To prove this claim, we only
consider the  case for $V^*$.
The argument for that of $V_{*}$ is similar.
 For each $i\in\M$, we want to show
\[
\disp
{\cal H}(s_0,x_0,y_0,i,V^*(s_0,x_0,y_0,i),D_s
\Phi(s_0,x_0,y_0), D_x \Phi(s_0,x_0,y_0),  D_y \Phi(s_0,x_0,y_0), D_{xx} \Phi(s_0,x_0,y_0)) \leq 0,
\]  for any test function $\Phi \in {\cal C}^{1,2,1}([0,T]\times \rr\times \rr_+) $
such that $(s_0,x_0,y_0, i)$ is a strictly local
 maximum of $V^* (s,x,y,i) -\Phi(s,x,y) $.
 Without loss of generality, we may assume
 that $V^*(s_0,x_0,y_0,i) = \Phi(s_0,x_0,y_0)$
 and because of the stability of our scheme we
 can also assume that
 $\Phi \geq  \sup_{k,h,l}\|V_{k,h,l}\|$ outside of the ball
 $ B((s_0,x_0,y_0),r) $
 where $r>0$ is such that
\[ \disp
V^*(s,x,y,i)- \Phi(s,x,y)\leq 0=V^*(s_0,x_0,y_0,i)-
\Phi(s_0,x_0,y_0) \ \hbox{ in } \  B((s_0,x_0,y_0),r).
\]
This implies that there exist sequences $k_n>0$,
$h_n >0$, $l_n>0$  and $(\theta_n, \eta_n,\zeta_n) \in[0,T]\times \rr\times\rr_+$
such that as $n\rightarrow \infty$ we have
\beq{limit3} \barray
\ad k_n \rightarrow 0, \,\,\,h_n\rightarrow 0,
\,\,\, l_n\to 0,\,\,\, \theta \to s_0,\,\,\,  \eta_n\rightarrow x_0,\, \,\,\zeta_n\rightarrow y_0,\,\,\,
V_{k_n,h_n,l_n}(\theta_n, \eta_n,\zeta_n,i)\rightarrow V^*(s_0,x_0,y_0,i),\, \nonumber \\
\ad \hbox{and}\,\,\, (\theta_n, \eta_n,\zeta_n) \,
\hbox{ is  a global maximum of } V_{k_n,h_n, l_n}-\Phi. \nonumber
\earray\eeq
Denote $\epsilon_n=V_{k_n,h_n, l_n}
(\theta_n,\eta_n,\zeta_n,i)-\Phi(\theta_n,\eta_n,\zeta_n)$. Obviously
$\epsilon_n \rightarrow 0$ and
\beq{limit4}V_{k_n,h_n,l_n}(s,x,y,i)\leq \Phi(s,x,y)
+ \epsilon_n  \ \hbox{ for all } \ (s,x,y) \in [0,T]\times  \rr\times \rr_+.
\eeq
We know that for all $\xi \in(0,1)$,
  \[
S(\xi,k_n,h_n,l_n,\theta_n,\eta_n,\zeta_n,i,V_{k_n,h_n, l_n}(\theta_n,\eta_n,\zeta_n,i),V_{k_n,h_n,l_n})=0
.\] The monotonicity of $S$ and (\ref{limit4}) implies
\beq{ineq3}  \barray \ad
S(\xi,k_n,h_n,l_n,\theta_n,\eta_n,\zeta_n,i,\Phi(\theta_n,\eta_n,\zeta_n) + \epsilon_n,\Phi + \epsilon_n)   \\
\aad  \ \le S(\xi,k_n,h_n,l_n,\theta_n,\eta_n,\zeta_n,i,V_{k_n,h_n, l_n}(\theta_n,\eta_n,\zeta_n,i),V_{k_n,h_n,l_n})=0.
\earray
\eeq
Therefore,
\[
\disp
\lim_{\xi\downarrow 0}\lim_{n\to \infty}S(\xi,k_n,h_n,l_n,\theta_n,\eta_n,\zeta_n,i,\Phi(\theta_n,\eta_n,\zeta_n) + \epsilon_n,\Phi + \epsilon_n)  \leq 0
,\]
so
\beq{ineq3000}\barray \ad
{\cal H}(s_0,x_0,y_0,i,V^*(s_0,x_0,y_0,i),D_s
\Phi(s_0,x_0,y_0), D_x \Phi(s_0,x_0,y_0),  D_y \Phi(s_0,x_0,y_0), D_{xx} \Phi(s_0,x_0,y_0))\leq 0. \nonumber
\earray \eeq
This proves that $V^*$ is a viscosity subsolution and,
similarly we can prove that $V_{*} $ is  a viscosity
supersolution. Thus, using the uniqueness of
the viscosity solution, we see that $V=V^*=V_{*}$. Therefore,
we
conclude that the sequence $(V_{h,k,l})_{h,k,l}$
converges  locally uniformly  to $V$ as desired.
\endproof

\section{Numerical Example}

In this example, we  present the optimal extraction and stopping policies of a mining company with an extraction contract with a 10 years maturity of an oil field with a known reserve of $Y(0) = 10$ billion barrels.  We assume that the oil  market has two main movements an uptrend and a downtrend. Thus the Markov chain $\al$ takes two states $ {\cal M}=\{1,2\}$ where $\al(t)=1$ denotes the uptrend and $\al(t)=2$ denotes the downtrend, the discount rate $r=0.005$, the return vector is $\mu=(0.01,-0.01)$, the volatility vector is $\sigma=(0.3,0.2)$, the intensity vector is $\gamma=(0.25,0.3)$, the time $T=10$ (in years), and the generator of the Markov chain is
\bea 
\disp
Q=\bigg(\begin{array}{ll} -0.003 & 0.003 \\
             0.005 &-0.005 
              \end{array} \bigg).
\eea
 We assume that the outputs may affect the market price of oil  so the drift of our diffusion has the form $\mu(t,x,u_1, i)  = x(\bar{\mu}(i)-\lambda u)$. The oil price dynamic is as follows
\bea
\disp \mathrm{d}X(t) =  X(t)\bigg((\bar{\mu}(\al(t))-\rho u_1(t))\mathrm{d}t +\sigma(\al(t))\mathrm{d}W_t + \int_\rr\gamma(\al(t)z\bar{N}(\mathrm{d}t, \mathrm{d}z)\bigg).
\eea
 The parameter $ \lambda\in[0,1)$ will capture the relative impact of  the oil production on the oil price, in this example $\lambda=0.001$
The running cost of the mine is $L(s,X(s),Y(s),u)=(X(s)-25)*u-5$ where the cost of extracting one barrel is \$25,  and the terminal cost $\Phi(s,X(s),Y(s))=(X(s)-30)Y(s)$, this captures the market value of the oil field at time s. We assume that the extraction rate $u(\cdot) \in  [0, 10000]$. Note that, the running and terminal costs are linear function of the extraction rate therefore using \thmref{Verification} the optimal extraction strategies will on be attained at the endpoints of the interval  $ [0, 5000]$, by looking at the sign of the derivative of the functional given in (\ref{differential_2}) with respect of $u$. The optimal extraction rate is given by
\bea
u^*(s) =\left\{\begin{array}{ll}\disp 5000 \quad \hbox{if }-\rho x \frac{\partial V(s,x,y,i)}{\partial x}-\frac{\partial V(s,x,y,i)}{\partial y}+ (x-25) >0\\
\disp 0\quad\hbox{if }-\rho x \frac{\partial V(s,x,y,i)}{\partial x}-\frac{\partial V(s,x,y,i)}{\partial y}+ (x-25) \leq 0.
\end{array} \right.
\eea
In Figures 1 and 2, the region above the curve represents when it is optimal to extract at full capacity $u^*(s)  =5000$ and the region below the curve represents when is optimal not to extract, $u^*(s) = 0$ when the market is up and when market is down. In Figures 3 and 4 represent the continuation region when the market is up and when the market is down. The region above the curve represents the continuation region $D$ defined in \thmref{Verification} where is optimal to keep extracting the minerals and the region below the curve represents the domain where it is preferable to shutdown the oil field.  

\begin{center}
\begin{figure}
\includegraphics[height=20cm,width=20cm]{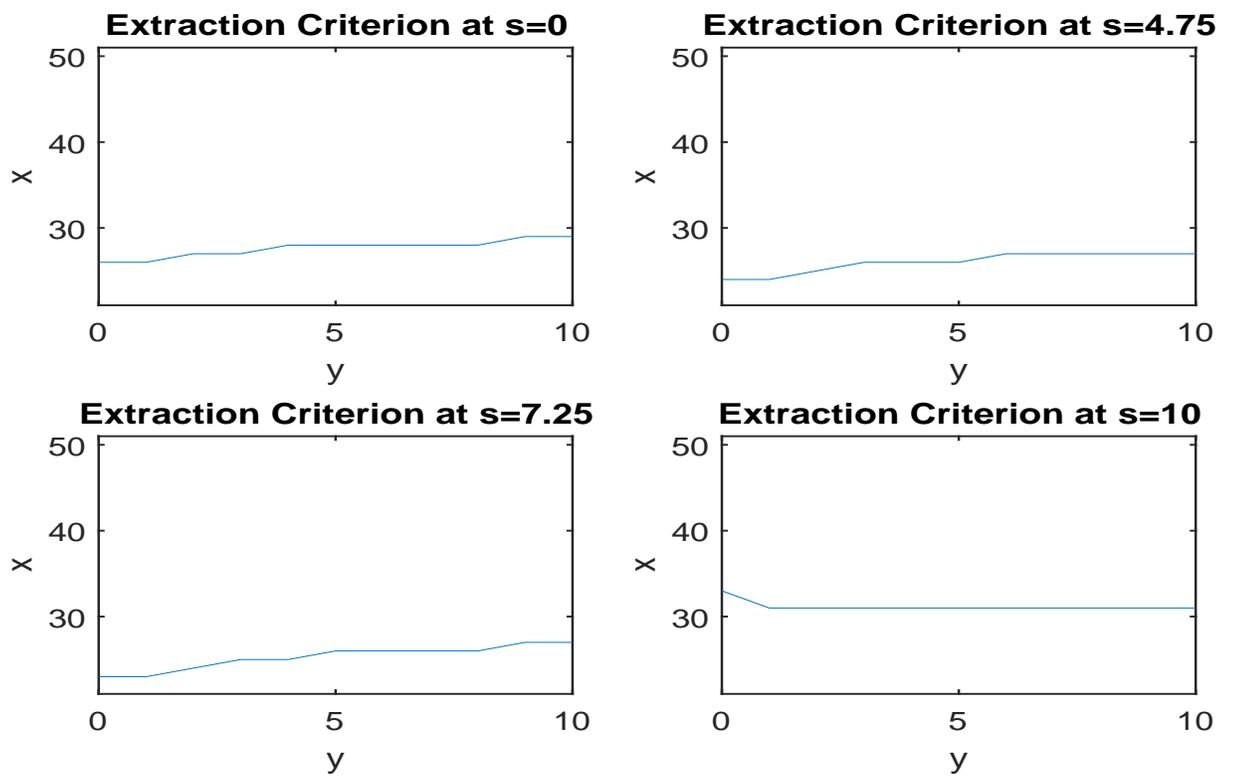}
\caption{Extraction criterion when the market is up at time $s=0, 4.75, 7.25, 10$. }
\end{figure}
\end{center}

\begin{center}
\begin{figure}
\includegraphics[height=20cm,width=20cm]{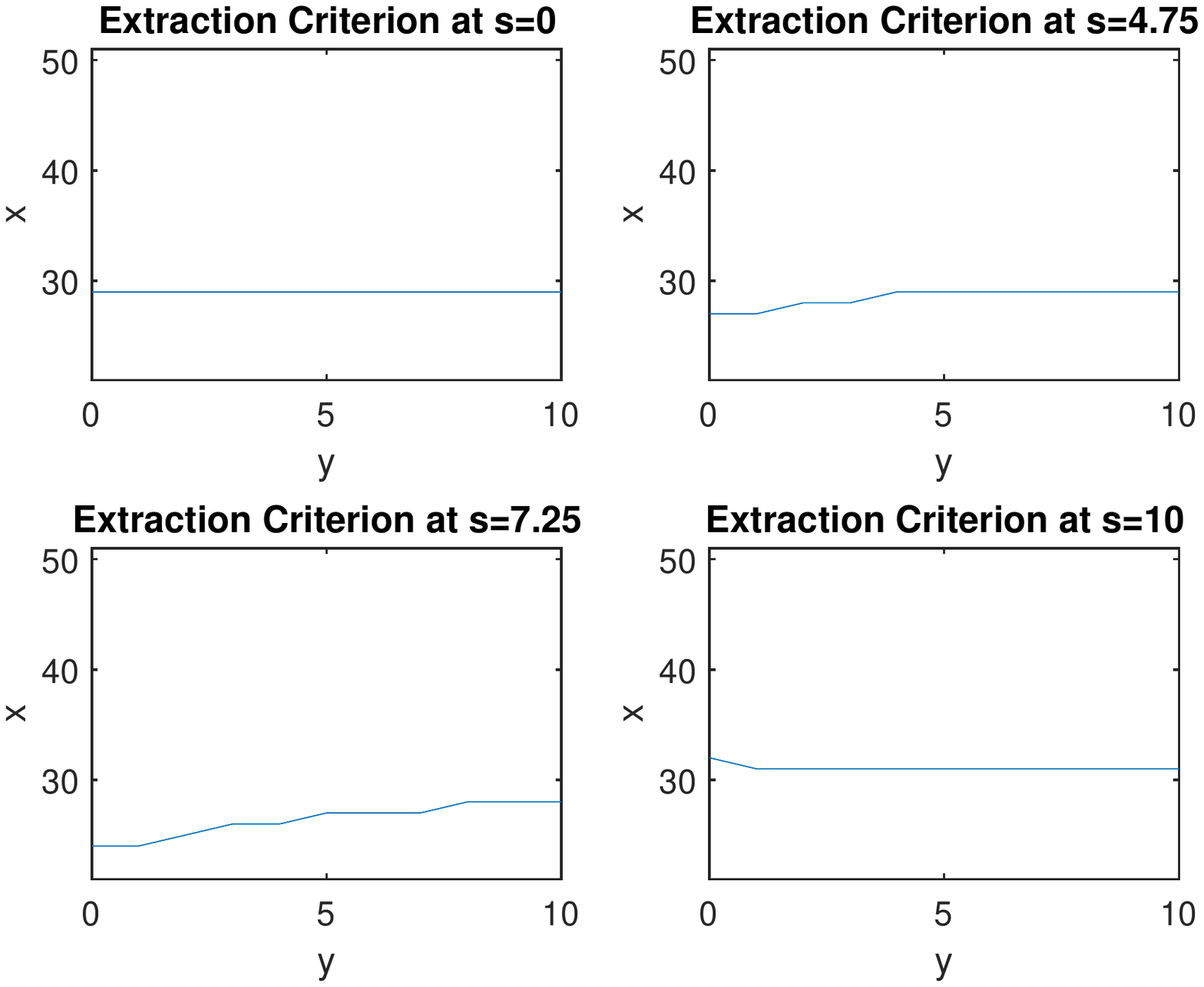}
\caption{Extraction criterion when the market is down at time $s=0, 4.75, 7.25, 10$. }
\end{figure}
\end{center}

\begin{center}
\begin{figure}
\includegraphics[height=20cm,width=20cm]{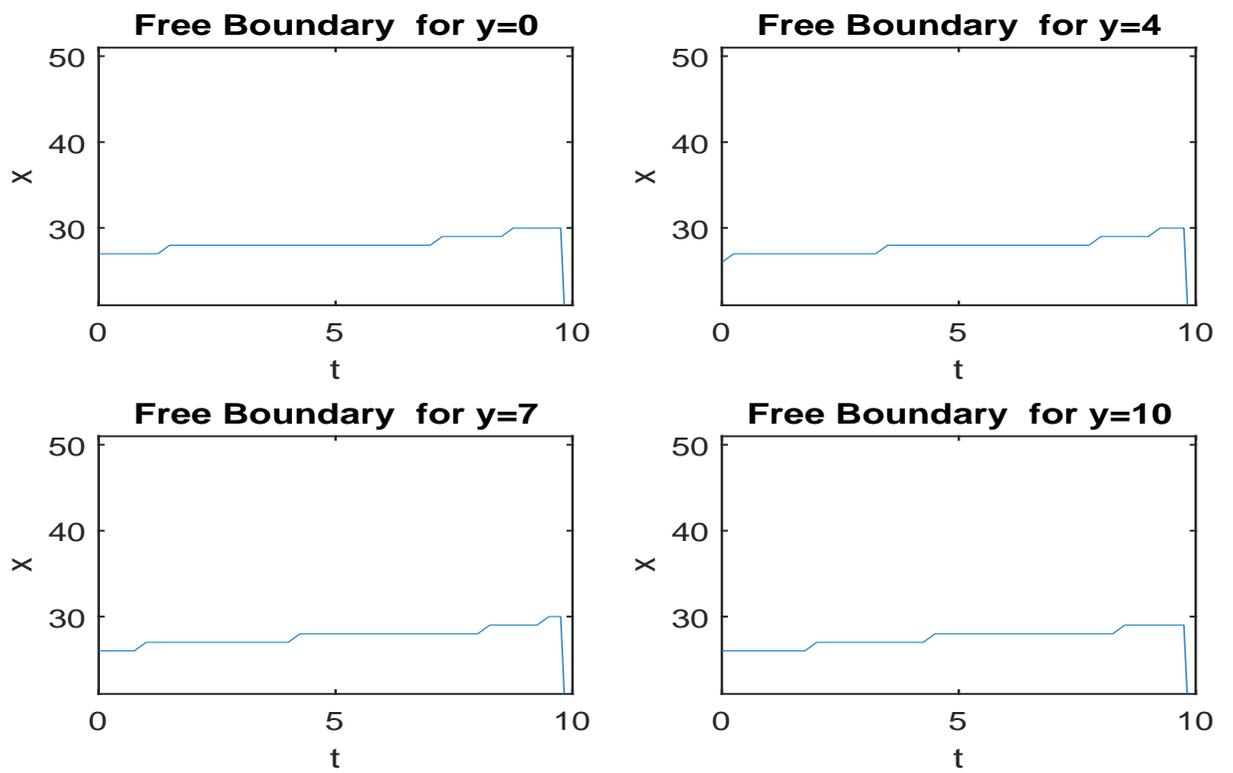}
\caption{Free boundary when the market is up for $y=0, 4, 7, 10$. }
\end{figure}
\end{center}

\begin{center}
\begin{figure}
\includegraphics[height=20cm,width=20cm]{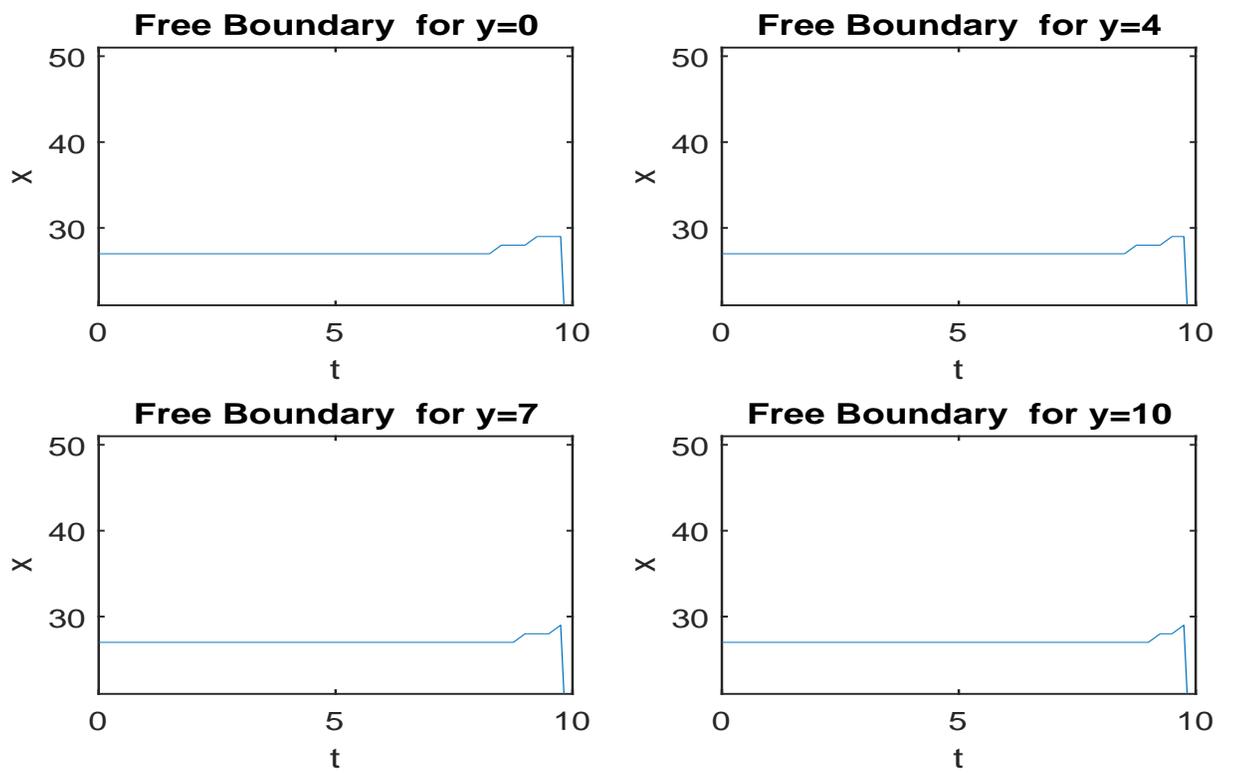}
\caption{Free boundary when the market is down for $y=0, 4, 7, 10$. }
\end{figure}
\end{center}

\end{document}